\documentclass[]{spie}  %>>> use for US letter paper
\usepackage[]{graphicx}

\title{Fractional charge in the noise of Luttinger liquid systems}

\author{Bj{\"o}rn Trauzettel\supit{a}, In{\`e}s Safi\supit{b},
Fabrizio Dolcini\supit{c}, and Hermann Grabert\supit{c}
\skiplinehalf \supit{a}Instituut-Lorentz, Universiteit Leiden,
2300 RA Leiden, The Netherlands; \\
\supit{b}Laboratoire de Physique des Solides, Universit\'e Paris-Sud,
91405 Orsay, France; \\
\supit{c}Physikalisches Institut, Albert-Ludwigs-Universit\"at, 79104
Freiburg, Germany
}

\authorinfo{Send correspondence to B. Trauzettel:
trauzettel@lorentz.leidenuniv.nl}
\begin{document}
  \maketitle

%%%%%%%%%%%%%%%%%%%%%%%%%%%%%%%%%%%%%%%%%%%%%%%%%%%%%%%%%%%%%
\begin{abstract}
The current noise of a voltage biased interacting quantum wire
adiabatically connected to metallic leads is computed in presence
of an impurity in the wire. We find that in the weak
backscattering limit the Fano factor characterizing the ratio
between shot noise and backscattering current crucially depends on
the noise frequency $\omega$ relative to the ballistic frequency
$v_F/gL$, where $v_F$ is the Fermi velocity, $g$ the Luttinger
liquid interaction parameter, and $L$ the length of the wire. In
contrast to chiral Luttinger liquids, the noise is not only due to
the Poissonian backscattering of fractionally charged
quasiparticles at the impurity, but also depends on Andreev-type
reflections of plasmons at the contacts, so that the frequency
dependence of the noise needs to be analyzed to extract the
fractional charge $e^*=e g$ of the bulk excitations. We show that
the frequencies needed to see interaction effects in the Fano
factor are within experimental reach.
\end{abstract}

%>>>> Include a list of keywords after the abstract

\keywords{Luttinger liquid, current noise, fractional charge}

%%%%%%%%%%%%%%%%%%%%%%%%%%%%%%%%%%%%%%%%%%%%%%%%%%%%%%%%%%%%%
\section{INTRODUCTION}
\label{sect:intro}  % \label{} allows reference to this section

Shot noise measurements are a powerful tool to observe the charge
of  elementary excitations of interacting electron systems. This
is due to the fact that in the Poissonian limit of uncorrelated
backscattering of quasiparticles from a weak impurity, the low
frequency current noise is directly proportional to the
backscattered charge \cite{blante00}. This property turns out to
be particularly useful in probing the fractional charge of
excitations in one-dimensional (1D) electronic systems, where
correlation effects destroy the Landau quasiparticle picture and
give rise to collective excitations, which in general obey
unconventional statistics, and which have a charge different from
the charge $e$ of an electron \cite{pham00}. In particular, for
fractional quantum Hall (FQH) edge state devices, which at filling
fraction $\nu=1/m$ ($m$ odd integer) are usually described by the
{\it chiral} Luttinger liquid (LL) model, it has been predicted
that shot noise should allow for an observation of the fractional
charge $e^*=e\nu$ of backscattered Laughlin quasiparticles
\cite{kane94}. Indeed, measurements at $\nu=1/3$ performed by two
groups at the Weizmann Institute in Israel \cite{depic97} and at
the CEA-Saclay in France \cite{samin97} have essentially confirmed
this picture.

The question arises whether similar results can be expected also
for {\it non-chiral} LLs, which are believed to be realized in carbon
nanotubes \cite{bockr99} and single channel semiconductor quantum
wires \cite{yacob96}. Although a non-chiral LL can be modelled
through the very same formalism as a pair of chiral LLs, some
important differences between these two kinds of LL systems have
to be emphasized. In particular, in chiral LL devices right- and
left-moving charge excitations are spatially separated, so that
their chemical potentials can be independently tuned in a
multi-terminal Hall bar geometry. In contrast, in non-chiral LL
systems, right- and left-movers are confined to the same channel,
and it is only possible to control the chemical potentials of the
Fermi liquid reservoirs attached to the 1D wire. This in turn
affects the chemical potentials of the right- and left-moving
charge excitations in a non-trivial way depending on the
interaction strength, and implies crucial differences between
chiral and non-chiral LLs, for instance, the conductance
in the former case depends on the LL parameter $g=\nu$ \cite{kane92},
while in the latter case it is independent of $g$
\cite{safi95,inhom95,safi97}. Hence, the
predictions on shot noise properties of FQH systems can not be
straightforwardly generalized to the case of non-chiral LLs,
which therefore deserve a specific investigation. Previous
theoretical calculations of the shot noise of non-chiral LL
systems have shown that, even in the weak backscattering limit,
the zero frequency noise of a finite-size non-chiral LL does not
contain any information about the fractional charge backscattered
off an impurity \cite{ponom99,trauz02}, but is rather proportional
to the charge of an electron. This result, as well as the above
mentioned interaction independent DC conductance, prevents easy
access to the interaction parameter $g$.

On the other hand, a quantum wire behaves as an Andreev-type
resonator for an incident electron, which is transmitted as
series of current spikes \cite{safi95}. The reflections of charge
excitations at both contacts are called Andreev-type reflections
because they are momentum conserving as ordinary Andreev
reflections \cite{safi95,sandl98}. Since the transmission dynamics
in the Andreev-type resonator depends on $g$, finite frequency
 transport can resolve internal properties of the wire.
This is, in fact, the case for the AC conductance
\cite{safi95,safi97,blant98}. However, finite frequency
conductance measurements are limited in the AC frequency range
since the frequency must be low enough to ensure quasi-equilibrium
states in the reservoirs in order to compare experiments to
existing theories. The better alternative is to apply a DC voltage
and measure finite frequency current noise. Exploring the out of
equilibrium regime, we have recently shown that the noise induced
by an impurity as a function of frequency has a periodic structure
with period $2\pi\omega_L$, where $\omega_L=v_F/gL$ is the inverse
of the traversal time of a charge excitation with plasmon velocity
$v_F/g$ through the wire of length $L$ \cite{trauz04}. The Fano
factor oscillates, and we demonstrate that by averaging over
$2\pi\omega_L$, the effective charge $e^*=e g$ can be extracted
from noise data. The interplay between the Andreev-type
reflections and the backscattering of charge excitations at the
impurity yields, in fact, many interesting phenomena both for the
DC current as well as the finite frequency noise. As far as the DC
current is concerned, the most remarkable feature is the
appearance of oscillations as a function of $eV/\hbar \omega_L$,
where $V$ is the applied voltage bias \cite{dolci03}. For the
finite frequency excess noise, we have found regions of negativity
showing interesting patterns which dramatically depend on the
electron-electron interaction strength in the wire and are
fundamentally different from their non-interacting counterparts
\cite{dolci05}.

\section{Inhomogeneous Luttinger liquid model}

\begin{figure}
   \begin{center}
   \begin{tabular}{c}
   \includegraphics[height=6cm]{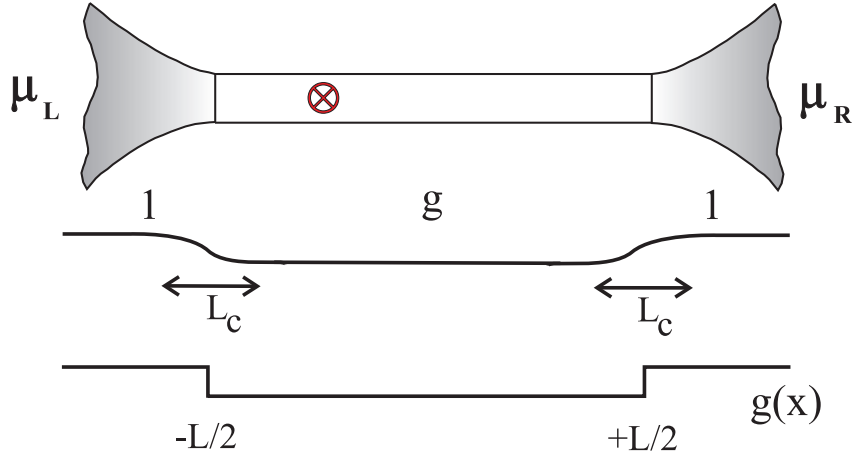}
   \end{tabular}
   \end{center}
   \caption[example]
%>>>> use \label inside caption to get Fig. number with \ref{}
   { \label{setup} The upper part of the figure shows a quantum
   wire with an impurity adiabatically coupled to Fermi liquid leads. In
order to allow for a finite bias, the leads are held on different
electro-chemical potentials $\mu_L$ and $\mu_R$. The middle part
of the figure shows the actual variation of the LL parameter $g$
along the wire-leads system in the inhomogeneous LL model, and the
lower part of the figure its simplification under the assumption
that $\lambda_F \ll L_c \ll L$.}
\end{figure}

In order to analyze the noise of non-chiral LLs, it is useful to
study the inhomogeneous LL model, schematically illustrated in
Fig.~\ref{setup}, which takes the finite length of the interacting
wire and the adiabatic coupling to the Fermi liquid leads
explicitly into account \cite{safi95,inhom95}. This model is
governed by the Hamiltonian
\begin{equation}
{\mathcal{H}} ={\mathcal{H}}_{0}  \, + \, {\mathcal{H}}_{B}  \, +
\, {\mathcal{H}}_{V} \; , \label{L}
\end{equation}
where ${\mathcal{H}}_{0}$ describes the interacting wire, the
leads and their mutual contacts, ${\mathcal{H}}_{B}$ accounts for
the electron-impurity interaction, and ${\mathcal{H}}_{V}$
represents the coupling to the electrochemical bias applied to
the wire. Explicitly, the three parts of the Hamiltonian read
\begin{eqnarray}
{\mathcal{H}}_0 &=&\frac{\hbar v_F}{2}  \int_{-\infty}^{\infty}
 dx \left[ \Pi^2 + \frac{1}{g^2(x)}
(\partial _x\Phi )^2\right]  \, , \label{L0}  \\
{\mathcal{H}}_B &=& \lambda \cos{[\sqrt{4 \pi} \Phi(x_0,t)+2 k_F
x_0]} \label{LB} \; ,\\
{\mathcal{H}}_{V}  &=&    \int_{-\infty}^{\infty}
\frac{dx}{\sqrt{\pi}} \, \mu(x) \,
\partial_x \Phi(x,t) \; . \label{LV}
\end{eqnarray}
Here, $\Phi(x,t)$ is the standard Bose field operator in
bosonization and $\Pi(x,t)$ its conjugate momentum density
\cite{gogol98}. The Hamiltonian ${\mathcal{H}}_0$ describes the
(spinless) inhomogeneous LL, which is known to capture the
essential physics of a quantum wire  adiabatically connected to
metallic leads. The interaction parameter $g(x)$ is
space-dependent and its value is 1 in the bulk of the
non-interacting leads
 and $g$ in the bulk of the wire ($0 < g < 1$ corresponding to repulsive
interactions). The variation of $g(x)$ at the contacts from 1 to
$g$ is assumed to be smooth, i.e. to occur within a characteristic
length $L_c$ fulfilling $\lambda_F \ll L_c \ll L$, where
$\lambda_F$ is the electron Fermi wavelength. The specific form of
the function $g(x)$ in the contact region is expected not to
influence physical properties for all energy scales below $\hbar
v_F/L_c$. A recent analysis performed on a lattice model version
of the inhomogeneous LL has confirmed this hypothesis
\cite{enss04}. Since the energy regime $E \geq \hbar v_F/L_c$ goes
beyond the interest of the present work \footnote{Transport
properties of a 1D interacting quantum dot, where energies of
order $\hbar v_F/L_c$ are important, have been analyzed in
Ref.~\citenum{kleim02}.}, we shall, henceforth, adopt a step-like
function, {\it cf.} Fig.~1.
%%%%%%%%%%%%%%%%%%%%%%%%%%%%%%%%%%%%%%%
%%%%%%%%%%%%%%%%%%%%%%%%%%%%%%%%%%%%%%%
%%%%%      FIGURE    1          %%%%%%
%%%%%%%%%%%%%%%%%%%%%%%%%%%%%%%%%%%%%%%
%%%%%%%%%%%%%%%%%%%%%%%%%%%%%%%%%%%%%%%

The Hamiltonian ${\mathcal{H}}_B$ is the dominant $2k_F$
backscattering term at the impurity site $x_0$, and introduces a
strong non-linearity in the field $\Phi$; the forward scattering
term caused by the impurity has been omitted since it does not
affect the statistics of the current. Finally, Eq.~(\ref{LV})
contains the applied voltage. In most experiments leads are normal
2D or 3D contacts, i.e.\ Fermi liquids. However, since we are
interested in properties of the wire, a detailed description of
the leads would in fact be superfluous. One can
 account for their main effect, the applied bias voltage at the
contacts, by treating them as non-interacting 1D systems ($g=1$),
as mentioned above. The only essential properties originating from
the Coulomb interaction that one needs to retain are (i) the
possibility to shift the band-bottom of the leads, and (ii)
electroneutrality \cite{trauz02}. Therefore, the function $\mu(x)$
appearing in Eq.~(\ref{LV}), which describes the externally
tunable electrochemical bias, is taken as piecewise constant
$\mu(x<-L/2) = \mu_L$, $\mu(x>L/2) = \mu_R$ corresponding to an
applied voltage $V = (\mu_L-\mu_R)/e$. In contrast, the QW itself
does not remain electroneutral in presence of an applied voltage,
and its electrostatics emerges naturally from
Eqs.~(\ref{L0})-(\ref{LV}) with $\mu=0$ for $|x|<L/2$
\cite{safi97,egger96}.

\section{Transport properties}

In this section, we discuss the features of the average current and
the current noise of the system depicted in Fig.~\ref{setup}.
The former reads
\begin{equation}
I = \langle j(x,t) \rangle ,
\end{equation}
where the current operator is related to the Bose
field $\Phi$ through $j (x,t) = - (e/\sqrt{\pi}) \partial_t \Phi(x,t)$.
Then, the latter, i.e.\ the finite frequency noise, is given by
\begin{eqnarray} \label{noise}
S(x,y;\omega) = \int_{-\infty}^{\infty} dt e^{i\omega t}
\left\langle \left\{ \Delta j (x,t) , \Delta j(y,0) \right\}
\right\rangle \; ,
\end{eqnarray}
where $\{ \, , \, \}$ denotes the anticommutator and $\Delta
j(x,t) = j(x,t) - \langle j(x,t) \rangle$ is the current
fluctuation operator. \footnote{The low frequency noise (with
$\omega x/v_F, \omega y/v_F \ll 1$) does not depend on the spatial
coordinates $x$ and $y$. This is, however, not true anymore at
high frequencies. \cite{trauz03}} Since we investigate
non-equilibrium properties of the system, the actual calculation
of the averages of current and noise are performed within the
Keldysh formalism \cite{keldy64}.

The average current $I$ can be expressed as $I=I_0 -I_{\rm BS}$,
where $I_0=(e^2/h) V$ is the current in the absence of an
impurity, and $I_{\rm BS}$ is the backscattering current. For
arbitrary impurity strength, temperature, and voltage, the
backscattering current can be written in the compact form
\begin{equation} \label{BScurrent}
I_{\rm BS}(x,t) = -\frac{\hbar \sqrt{\pi}}{e^2}
\int_{-\infty}^\infty dt' \sigma_0(x,t;x_0,t') \langle j_B
(x_0,t') \rangle_\rightarrow \ ,
\end{equation}
where $\sigma_0(x,t;x_0,t')$ is the non-local conductivity of the
clean wire \cite{safi95,safi97,blant98}. In Eq.~(\ref{BScurrent}),
we have introduced the ``backscattering current operator''
\begin{equation} \label{BSoperator}
j_B(x_0,t) \equiv - \frac{e}{\hbar} \frac{\delta
{\mathcal{H}}_B}{\delta \Phi(x_0,t)} (\Phi + A_0) \ ,
\end{equation}
where $A_0(x_0,t)$ is a shift of the phase field emerging when one
gauges away the applied voltage. For a DC voltage this shift
simply reads $A_0(x_0,t)=\omega_0 t/2\sqrt{\pi}$ with $\omega_0 =
eV/\hbar$ and $I_{\rm BS}$ does not depend on $x$ and $t$.
Furthermore, we have introduced a ``shifted average''
$\langle \dots \rangle_\rightarrow$, which is evaluated with
respect to the shifted Hamiltonian
\begin{equation} \label{shift}
{\mathcal H}_\rightarrow = {\mathcal{H}}_0[\Phi] + {\mathcal{H}}_B[\Phi+A_0]
\; .
\end{equation}
A straightforward though
lengthy calculation shows that the finite frequency current noise
(\ref{noise}) can (again for arbitrary impurity strength,
temperature, and voltage) be written as the sum of three contributions
\begin{equation} \label{ff_result}
S(x,y;\omega)=S_0(x,y;\omega)+S_A(x,y;\omega)+S_C(x,y;\omega) \; .
\end{equation}
The first part of Eq.~(\ref{ff_result}), $S_0(x,y;\omega)$, is the
current noise in the absence of a backscatterer, and can be
related to the conductivity $\sigma_0(x,y;\omega)$ by the
fluctuation dissipation theorem \cite{ponom96}
\begin{equation}
S_0(x,y;\omega)= 2 \hbar \omega \coth\left( \frac{\hbar \omega}{2 k_B T}
\right) \Re [\sigma_0(x,y;\omega)] \; .
\label{s0thermal}
\end{equation}
The conductivity can be expressed by the Kubo formula
$\sigma_0(x,y;\omega) =  2 (e^2/h) \omega C_0^R(x,y;\omega)$, where
\begin{equation}
C_0^R(x,y;\omega) =  \int_0^\infty dt e^{i \omega t} \langle [
\Phi(x,t), \Phi(y,0)] \rangle_0
\end{equation}
is the time-retarded correlator of the inhomogeneous LL model in
the absence of an impurity. It is important to note that usually
the relation (\ref{s0thermal}) is only valid in thermal
equilibrium, and the Kubo formula is based on linear response
theory. However, due to the fact that in the absence of an
impurity the current of a quantum wire attached to Fermi liquid
reservoirs is linear in the applied voltage \cite{safi95,safi97},
Eq.~(\ref{s0thermal}) is also valid out of equilibrium.

The other two terms in Eq.~(\ref{ff_result}) arise from the
partitioning of the current at the impurity site and reflect the
statistics of the backscattering current. The second term is
related to the anticommutator of the backscattering current
operator $j_B$, and reads
\begin{eqnarray} \label{sb}
&& S_A(x,y;\omega) = \frac{1}{\pi} \left(\frac{h}{2 e^2}\right)^2
\sigma_0(x,x_0;\omega) f_A(x_0,\omega) \sigma_0(x_0,y;-\omega)
\end{eqnarray}
with
\begin{equation}
f_A(x_0,\omega) = \int_{-\infty}^\infty dt \, e^{i \omega t}
\left\langle \left\{ \Delta j_B(x_0,t), \Delta j_B(x_0,0)
\right\} \right\rangle_\rightarrow \; ,
\end{equation}
where $\Delta j_B(x,t) \equiv j_B(x,t) - \langle j_B(x,t) \rangle_\rightarrow$.
Finally, the third part of Eq.~(\ref{ff_result})
is related to the time-retarded commutator of $j_B$ and can
be expressed as
\begin{eqnarray} \label{sm2}
S_C(x,y;\omega) = \frac{h}{2 e^4 \omega} \Bigl\{ S_0(x,x_0;\omega)
f_C(x_0,-\omega) \sigma_0(x_0,y;-\omega) - S_0(y,x_0;-\omega)
f_C(x_0,\omega) \sigma_0(x_0,x;\omega) \Bigr\}
\end{eqnarray}
with
\begin{equation}
f_C(x_0,\omega) = \int_0^\infty dt \left( e^{i \omega t}-1 \right)
\left\langle \left[ j_B(x_0,t),j_B(x_0,0) \right]
\right\rangle_\rightarrow \; .
\end{equation}
The fractional charge is expected to emerge only in the limit of
weak backscattering through the ratio between shot noise and
backscattering current. We thus focus on the case of a weak
impurity, retaining in the expressions (\ref{BScurrent}) and
(\ref{ff_result}) only contributions of second order in the
impurity strength $\lambda$. Furthermore, we concentrate on the
shot noise limit of large applied voltage $eV \gg \{ k_B T,
\hbar\omega, \hbar \omega_L \}$. Properties of the equilibrium
current noise at finite frequency and temperature are discussed in
detail in Ref.~\citenum{dolci05}.

\section{Fano factor}

The backscattering current (\ref{BScurrent}) may be written as
$I_{\rm BS} = (e^2/h) {\cal R} V $, where ${\cal R}$ is an
effective reflection coefficient. Contrary to a non-interacting
electron system, ${\cal R}$ depends   on voltage and interaction
strength \cite{kane92,dolci03}. In the weak backscattering limit
${\cal R} \ll 1$, and its actual value can readily be determined
from a measurement of the current voltage characteristics.
Importantly, in the shot noise regime, i.e.\ for $eV \gg \{ k_B T,
\hbar\omega, \hbar \omega_L \}$, the
noise can be shown to be dominated by the second term in
Eq.(\ref{ff_result}) and to take the simple form \cite{dolci05}
\begin{equation} \label{excessnoise}
S (x,x;\omega) \simeq  2 e F(\omega) I_{\rm BS} \; ,
\end{equation}
where $x=y$ is the point of measurement (in either of the two
leads). The Fano factor
\begin{equation}
F(\omega)= \frac{h^2}{e^4} |\sigma_0(x,x_0;\omega)|^2
\label{Fano-cond}
\end{equation}
is given in terms of the non-local conductivity
$\sigma_0(x,x_0;\omega)$ relating the measurement point $x$ to the
impurity position $x_0$, and reads explicitly
\begin{eqnarray} \label{Fano-fun}
F(\omega) =   ( 1-\gamma)^2 \frac{1+\gamma^2+2\gamma \cos \left(
\frac{2\omega (\xi_0+1/2)}{\omega_L} \right)}{1+\gamma^4-2\gamma^2
\cos \left( \frac{2 \omega}{\omega_L}\right)} \; .
\end{eqnarray}
The latter expression is, in fact, independent of the point of
measurement $x$ and of temperature. On the other hand, it depends,
apart from the frequency $\omega$, on the (relative) impurity
position $\xi_0=x_0/L$, and the interaction strength through
$\gamma = (1-g)/(1+g)$.
%%%%%%%%%%%%%%%%%%%%%%%%%%%%%%%%%%%%%%%
%%%%%%%%%%%%%%%%%%%%%%%%%%%%%%%%%%%%%%%
%%%%%      FIGURE    2          %%%%%%
%%%%%%%%%%%%%%%%%%%%%%%%%%%%%%%%%%%%%%%
%%%%%%%%%%%%%%%%%%%%%%%%%%%%%%%%%%%%%%%
\begin{figure}
   \begin{center}
   \begin{tabular}{c}
   \includegraphics[height=8cm]{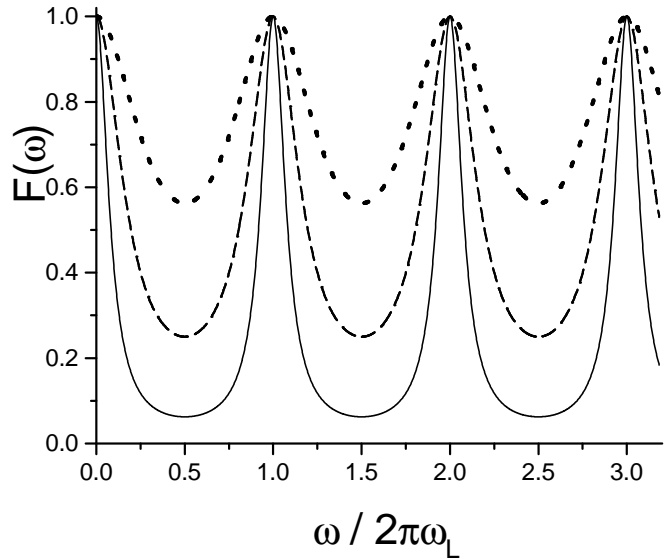}
   \end{tabular}
   \end{center}
   \caption[example]
%>>>> use \label inside caption to get Fig. number with \ref{}
   { \label{Fano_x000} The periodic function $F(\omega)$,
which determines the Fano factor, is shown as a function of
$\omega/2 \pi\omega_L$, for the case of an impurity at the center
of the wire ($x_0=0$) and three different values of the
interaction strength: $g=0.25$ (solid), $g=0.50$ (dashed), and
$g=0.75$ (dotted). In the regime $\omega/\omega_L \ll 1$, the
function tends to 1 independent of the value of $g$, but for
$\omega \approx \omega_L$ the curve strongly depends on the
interaction parameter $g$. In particular, $g$ can be obtained as
the average over one period.}
\end{figure}

The result (\ref{excessnoise}) shows that the ratio between the
shot noise and the backscattering current crucially depends on the
frequency regime one explores. In particular, for $\omega
\rightarrow 0$, the function $F$ tends to 1, independent of the
value of the interaction strength. Therefore, in the regime
$\omega \ll \omega_L$ the observed charge is just the electron
charge. In contrast, at frequencies comparable to $\omega_L$ the
behavior of $F$ as a function of $\omega$ strongly depends on the
LL interaction parameter $g$, and signatures of LL physics emerge.
This is shown in Fig.~\ref{Fano_x000} for the case of an impurity
located at the center of the wire. Then, $F(\omega)$ is periodic,
and the value at the minima coincides with $g^2$. Importantly, $g$
is also the mean value of $F$ averaged over one period $2\pi
\omega_L$,
\begin{equation} \label{s_ex_average}
\left\langle S (x,x;\omega) \right\rangle_{\omega} \equiv
\frac{1}{2 \pi \omega_L} \int_{-\pi \omega_L}^{\pi \omega_L}
S (x,x;\omega) \simeq 2 e g I_{\rm BS} \; ,
\end{equation}
where the approximation (\ref{excessnoise}) has been used to
describe $S(x,x;\omega)$. Seemingly, Eq.~(\ref{s_ex_average})
suggests that quasiparticles with a fractional charge $e^*=eg$ are
backscattered off the impurity in the quantum wire.

Let us discuss the physical origin of this appearance of
fractional charge: We first consider the case of an infinitely
long quantum wire. In the limit $L\to\infty$, i.e.\ $\omega_L\to
0$, $\xi_0 \rightarrow 0$, the function $F(\omega)$ becomes
rapidly oscillating and its average over any finite frequency
interval approaches $g$. Hence, we recover in this limit the
result for the homogeneous LL system \cite{kane94}, where the shot
noise is directly proportional to the fractional charge $e^*=ge$
backscattered off the impurity. However, as shown above, the value
of the fractional charge $e^*$ can be extracted not only in the
borderline case $\omega\gg\omega_L$, but already for frequencies
$\omega$ of order $\omega_L$. This is due to the fact that,
although the contacts are adiabatic, the mismatch between
electronic excitations in the leads and in the wire inhibits the
direct penetration of electrons from the leads into the wire;
rather a current pulse is decomposed into a sequence of fragments
by means of Andreev-type reflections at the contacts
\cite{safi95}. These reflections are governed by the coefficient
$\gamma=(1-g)/(1+g)$, which depends on the interaction strength.
The zero frequency noise is only sensitive to the sum of all
current fragments, which add up to the initial current pulse
carrying the charge $e$. However, when $2\pi/\omega$ becomes
comparable to the time needed by a plasmonic excitations to travel
from the contact to the impurity site, the noise resolves the
current fragmentation at the contacts. The sequence of
Andreev-type processes is encoded in the non-local conductivity
$\sigma_0(x,x_0;\omega)$ relating the measurement point $x$ and
the impurity position $x_0$. This enters into the Fano factor
(\ref{Fano-cond}) and allows for an identification of $e^*$ from
finite frequency noise data.
%%%%%%%%%%%%%%%%%%%%%%%%%%%%%%%%%%%%%%%
%%%%%%%%%%%%%%%%%%%%%%%%%%%%%%%%%%%%%%%
%%%%%      FIGURE    3          %%%%%%
%%%%%%%%%%%%%%%%%%%%%%%%%%%%%%%%%%%%%%%
%%%%%%%%%%%%%%%%%%%%%%%%%%%%%%%%%%%%%%%
\begin{figure}
   \begin{center}
   \begin{tabular}{c}
   \includegraphics[height=8cm]{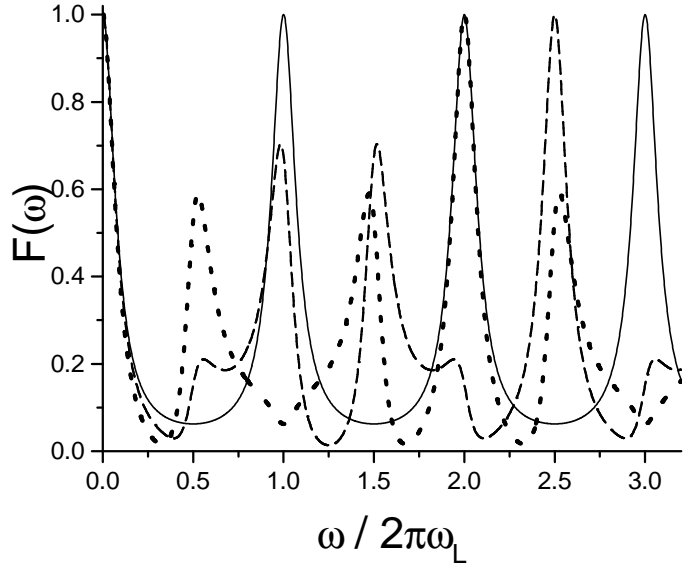}
   \end{tabular}
   \end{center}
   \caption[example]
%>>>> use \label inside caption to get Fig. number with \ref{}
   { \label{Fano_g025}  The Fano factor $F(\omega)$ is shown for
the interaction strength $g=0.25$ and three different values of
the (relative) impurity position $\xi_0=x_0/L$: $\xi_0=0$ (solid),
$\xi_0=0.10$ (dashed), and $\xi_0=0.25$ (dotted). The combined
effect of strong interactions and an off-centered impurity yields
a very pronounced suppression of the Fano factor for certain noise
frequencies.}
\end{figure}

\begin{figure}
   \begin{center}
   \begin{tabular}{c}
   \includegraphics[height=10cm]{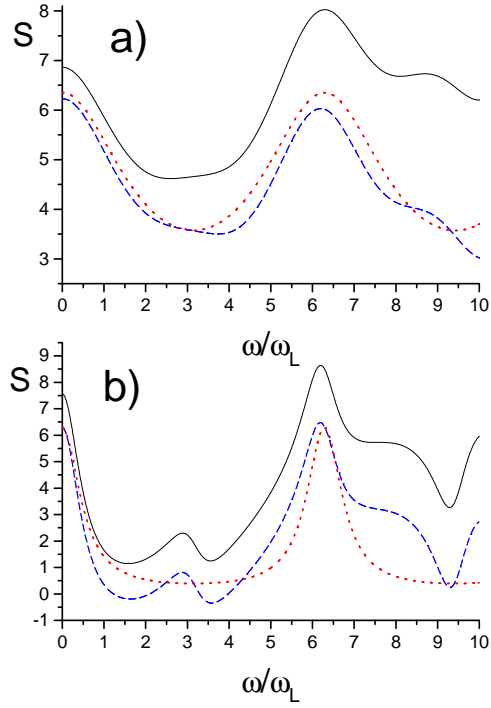}
   \end{tabular}
   \end{center}
   \caption[example]
{\label{full_noise1} The frequency spectrum of the full noise
$S(x,x;\omega)$, (black) solid line, in units of $e^2 \omega_L$
and the approximation (\ref{excessnoise}), (red) dotted line, are
depicted for two systems with the parameters a) $g=0.75$, $k_B
T/\hbar \omega_L=1$, $eV/\hbar \omega_L=100$, $\mathcal{R}=0.2$,
and b) $g=0.25$, $k_B T/\hbar \omega_L=2$, $eV/\hbar
\omega_L=98.83$, $\mathcal{R}=0.2$. The impurity is chosen at the
center of the wire. The (blue) dashed curve resembles the full
noise after a background subtraction, i.e.\
$S(x,x;\omega)-(e^2/\pi) \omega \coth ( \hbar \omega/2k_B T )$.}
\end{figure}

When the impurity is located away from the center of the wire,
$F(\omega)$ is no longer strictly periodic, as shown in Fig.
\ref{Fano_g025}. In that case, the combined effect of Coulomb
interactions and an off-centered impurity can lead to a very
pronounced reduction of the Fano factor for certain noise
frequencies (see Fig.~\ref{Fano_g025}). Moreover, even if the
impurity is off-centered, the detailed predictions
(\ref{excessnoise}) and (\ref{Fano-fun}) should allow to gain
valuable information on the interaction constant $g$ from the low
frequency behavior of the Fano factor determined by
\begin{equation}
F(\omega) = 1-(1-g^2) \Bigl( 1+4 g^2 \xi_0(1+\xi_0) \Bigr)
\left(\frac{\omega L}{2 v_F}\right)^2 + {\cal O}
\left(\frac{\omega}{\omega_L}\right)^4 .
\end{equation}
Finally, we would like to specify in more detail the parameter range
for the validity of the relation (\ref{excessnoise}).
It turns out that the optimal
voltage and temperature regime to extract (\ref{excessnoise}) from the
expression (\ref{ff_result}) for the full noise depends on the strength of
the electronic correlations in the wire \cite{dolci05}. In
Fig.~\ref{full_noise1}, we illustrate two characteristic cases,
namely $g=0.75$ (as appropriate for semiconductor quantum wires)
and $g=0.25$ (as appropriate for metallic single-wall carbon
nanotubes). The figure shows that the shot noise approximation
(\ref{excessnoise}) underestimates the full noise
(\ref{ff_result}) even at large applied voltages. However,
once a background noise
\begin{equation}
S_{\rm BG} = \frac{e^2}{\pi} \omega \coth \left( \frac{\hbar
\omega}{2k_B T} \right)
\end{equation}
has been subtracted from the full noise, which is done in order to
obtain the (blue) dashed curve in Fig.~\ref{full_noise1}, the
approximation (\ref{excessnoise}) and $S(x,x;\omega)-S_{\rm BG}$
agree within a few percent. The agreement becomes better the
weaker the interactions. This is due to the fact that, in the weak
interaction regime, the reflection coefficient of the Andreev-type processes
is rather small and, therefore, the
interference patterns in transport properties of the system are
less pronounced.
%%%%%%%%%%%%%%%%%%%%%%%%%%%%%%%%%%%%%%%
%%%%%%%%%%%%%%%%%%%%%%%%%%%%%%%%%%%%%%%
%%%%%        FIGURE   4          %%%%%%
%%%%%%%%%%%%%%%%%%%%%%%%%%%%%%%%%%%%%%%
%%%%%%%%%%%%%%%%%%%%%%%%%%%%%%%%%%%%%%%

\section{Conclusions}

The finite frequency noise of an interacting quantum wire coupled
to Fermi liquid leads has been analyzed using the Keldysh
technique for the inhomogeneous Luttinger liquid model. It has
been shown that the interplay between pure interaction effects of
the wire itself and the boundary effects at the crossover points
between the wire and the leads results in non-trivial features of the
 transport properties of the system. The appearance of fractional charge
$e^*=eg$ in the finite frequency noise of non-chiral LLs is solely
due to a combined effect of backscattering of bulk quasiparticles
at the impurity and of Andreev-type reflections of plasmons at the
interfaces of wire and leads. The fractional charge $e^*$
 can be extracted from the noise by
averaging it over a frequency range of size $\pi\omega_L$ in the
out of equilibrium regime. For single-wall carbon nanotubes we
know that $g \approx 0.25$, $v_F \approx 10^5 \; {\rm m/s}$, and
their length can be up to 10 microns. Thus, we estimate
$\pi\omega_L \approx 100 \; {\rm GHz} \dots 1 \; {\rm THz}$, which
is a frequency range that seems to be experimentally accessible
\cite{schoe97,deblo03}. Moreover, the requirement $eV \gg \hbar
\omega_L$ should be fulfilled in such systems for $eV \approx 10
\dots 50 \; {\rm meV}$, a value which is well below the subband
energy separation of about $1 \; {\rm eV}$. Finally, we mention
that similar oscillations as the ones predicted by us for the Fano
factor of the finite frequency noise of a quantum wire with an
impurity have recently been reported for the noise of a setup in
which an STM tip allows for electron tunneling into a clean
interacting quantum wire. \cite{marti04}

\acknowledgments

We thank F.~Ballestro, H.~Bouchiat, R.~Deblock, R.~Egger,
D.~C.~Glattli, L.~P.~Kouwenhoven, E.~Onac, and P. Roche for
interesting discussions. Financial support by EU Training Networks
(DIENOW and SPINTRONICS) and by the Dutch Science Foundation
NWO/FOM is gratefully acknowledged.

%%%%%%%%%%%%%%%%%%%%%%%%%%%%%%%%%%%%%%%%%%%%%%%%%%%%%%%%%%%%%
%%%%% References %%%%%

\bibliography{report}   %>>>> bibliography data in report.bib
\bibliographystyle{spiebib}   %>>>> makes bibtex use spiebib.bst

\end{document}